\documentclass[a4paper,fleqn,usenatbib,useAMS]{mnras}

\usepackage{graphicx}   
\usepackage{amsmath}    
\usepackage{amssymb}    
\usepackage{multicol}   
\usepackage{bm}         
\usepackage{pdflscape}  

\def\la{\;
\raise0.3ex\hbox{$<$\kern-0.75em\raise-1.1ex\hbox{$\sim$}}\; }
\def\ga{\;
\raise0.3ex\hbox{$>$\kern-0.75em\raise-1.1ex\hbox{$\sim$}}\; }

\newcommand{\daa}{$\Delta\alpha/\alpha$}
\newcommand{\dmm}{$\Delta\mu/\mu$}
\newcommand{\dmmt}{$\Delta\tilde{\mu}/\tilde{\mu}$}
\newcommand{\dV}{$\Delta V$}

\newcommand{\dFF}{$\Delta F/F$}
\newcommand{\kms}{km~s$^{-1}$}
\newcommand{\ms}{m~s$^{-1}$}
\newcommand{\etal}{{et al.}}

\newcommand{\CI}{[C~{\sc i}]}
\newcommand{\CII}{[C~{\sc ii}]}

\newcommand{\MgII}{Mg~{\sc ii}}

\usepackage[T1]{fontenc}
\usepackage{ae,aecompl}

\title[Limits on $\mu$-variation at $z > 6$]
{\textit{Constraints on the electron-to-proton mass ratio variation
at the epoch of reionization
}}

\author[S. A. Levshakov \etal ] {
S. A. Levshakov$^{1,2,3}$\thanks{E-mail: lev@astro.ioffe.ru},
M. G. Kozlov$^{2,3}$,
I. I. Agafonova$^{3}$
\vspace*{8pt}
\\
$^{1}$Ioffe Physical-Technical Institute, 194021 St.~Petersburg, Russia\\
$^{2}$Petersburg Nuclear Physics Institute, 188300 Gatchina, Russia\\
$^{3}$Electrotechnical University ``LETI'', 197376 St.~Petersburg, Russia
}
\date{Accepted 2020 August 25. Received 2020 August 25; in original form 2020 July 29}

\pubyear{2020}

\begin{document}
\label{firstpage}
\pagerange{\pageref{firstpage}--\pageref{lastpage}}
\maketitle

\begin{abstract}
Far infrared fine-structure transitions of \CI\ and \CII\ and rotational transitions of CO
are used to probe hypothetical variations of the electron-to-proton mass ratio 
$\mu = m_{\rm e}/m_{\rm p}$ at the epoch of reionization ($z > 6$).
A constraint on 
\dmm~= $(\mu_{\rm obs} - \mu_{\rm lab})/\mu_{\rm lab} = (0.7 \pm 1.2)\times10^{-5}$
($1\sigma$) obtained at $\bar{z} = 6.31$ is the most stringent up-to-date limit on 
the variation of $\mu$ at such high redshift. For all available estimates of \dmm\ 
ranging between $z = 0$ and $z \sim 1100$,~-- the epoch of recombination,~--
a regression curve \dmm~= $k_\mu (1+z)^p$, with $k_\mu = (1.6 \pm 0.3)\times10^{-8}$
and $p = 2.00 \pm 0.03$, is deduced. If confirmed, this would imply a dynamical 
nature of dark matter/dark energy. 
\end{abstract}



\section{Introduction}
\label{Sec1}

A plethora of models for the dark sector (dark matter and dark energy)
suppose the existence of Higgs-like scalar field(s) which couple non-universally to 
the matter content of the Standard Model (SM) of particle physics
(for reviews see, e.g., Battaglieri \etal\ 2017; Irastorza \& Redondo 2018; Beacham \etal\ 2019).
Such coupling could change the masses of particles, thus leading to violation
of the weak equivalence principle, and giving rise to the so-called 5$th$ force.
However, all attempts to discover traces of any interaction beyond the SM 
in experiments on Earth and even in satellite missions have led to a null result 
(Thompson 2019a,b, 2018, 2017; Banerjee \etal\ 2018; Berg\'e \etal\ 2018; Antoniou \& Perivolaropoulos 2017;
Rider \etal\ 2016; Brax \& Davis 2016; Li \etal\ 2016; Hamilton \etal\ 2015; Wagner \etal\ 2012).
This means that the 5$th$ force~-- if exists~-- should be extremely long-ranged 
(with characteristic scales of order of galactic to intergalactic distances) and/or some
screening mechanisms should be present which suppress the coupling strength in the
environments where the experiments were performed (Brax 2018).
In this respect the objects where the non-standard interactions could be expected from either 
the phenomenology or from theoretical considerations are more amenable to search for hidden fields. 
 
The rotational curve of the Milky Way (MW) shows that the total matter (and the gravitational potential) 
within the solar circle is dominated by baryons with 
$\rho_{\rm b}/\rho_{\scriptscriptstyle{\rm DM}} \sim 10$,
where $\rho_{\rm b}$ is the baryon density and $\rho_{\scriptscriptstyle{\rm DM}}$ is the density of 
dark matter (DM) (McGaugh 2018; McMillan 2017; Iocco \etal\ 2015; Sofue \etal\ 2009).
Thus, the non-detection of a signal from the dark sector in
objects located in the vicinity of the galactic position of the Sun
is at least conceivable. On the other hand, studies of stellar dynamics revealed
the regions where the gravitational potential indeed becomes DM-dominated~-- these are the low surface
brightness dwarf galaxies or the outskirts of 'normal' galaxies. 
In the quest for the 'dark' signal, targets in such regions seem to be more favorable.  

Extremely high-redshift objects also represent a perspective group. 
For instance, investigations of the ionization processes at redshift $z \sim 1100$
which are responsible for the temperature and polarization anisotropies of the
cosmic microwave background (CMB) radiation (Hannestad 1999; Kaplinghat \etal\ 1999; Kujat \& Scherrer 2000;
Yoo \& Scherrer 2003; Ichikawa \etal\ 2006; Planck Collaboration \etal\ 2015; Hart \& Chluba 2020)
show that the value of the Hubble constant $H_0$ correlates with the electron mass.
As a consequence, an increased effective electron mass at the epoch of recombination,
$m_{{\rm e},z} = (1.0190\pm0.0055)m_{\rm e,0}$, leads to a shifted value of the Hubble constant 
$H_0 \simeq 71$ km~s$^{-1}$~Mpc$^{-1}$ as compared with that inferred from the $\Lambda$ cold dark 
matter model ($\Lambda$CDM) calibrated by Planck CMB data $H_0 = 67.4\pm0.5$ km~s$^{-1}$~Mpc$^{-1}$
(Planck Collaboration \etal\ 2018).  This interplay between $m_{{\rm e},z}$ and $H_0$ may alleviate 
the so-called Hubble tension~-- the difference between the Hubble constant measured in the late 
Universe ($z \la 1$), $H_0 = 74.03\pm1.42$ km~s$^{-1}$~Mpc$^{-1}$ (Riess \etal\ 2019) and
$H_0 = 73.3\pm1.8$ km~s$^{-1}$~Mpc$^{-1}$ (Wong \etal\ 2019), and the CMB value.

Another high-redshift control point is the cosmic dawn ($z \la 20$). The depth of the H~{\sc i} 21-cm
absorption trough at $z = 14-21$ detected in EDGES by Bowman \etal\ (2018) turned out to be twice as large
as predicted what required additional mechanisms to explain it. Many models consider now different types 
of non-gravitational interaction between dark and baryonic matter as factors which could cool
the neutral gas (Barkana \etal\ 2018; Houston \etal\ 2018; Fraser \etal\ 2018; Safarzadeh \etal\ 2018; 
Yang W. \etal\ 2019; Famaey \etal\ 2020).

The subsequent epoch of reionization ($6 < z < 10$) is characterized by the progressive ionization of
the previously neutral intergalactic medium that makes possible observations across much of the electromagnetic 
spectrum. This results in a variety of methods which can be devised to test theoretical models beyond the SM. 
In particular, in the present paper we probe the non-standard coupling in two objects at $z > 6$
by means of radio spectroscopy of atomic and molecular transitions. Note that the cosmic time interval 
between $z=6$ and $z=1100$ is about 900 Myr, i.e., less than 1/10$th$ of the Hubble time, and between 
$z=6$ and $z=17$ is only 640 Myr\footnote{Here we adopt a flat $\Lambda$CDM cosmology with 
$H_0 = 70$ km~s$^{-1}$~Mpc$^{-1}$, $\Omega_\Lambda = 0.7$, $\Omega_{\rm m} = 0.3$.}.  

The theoretically predicted coupling affects preferentially the electron mass, whereas the
mass of the proton is determined by the strength of the strong interaction of quarks 
and remains less affected. Therefore, the electron-to-proton mass ratio $\mu = m_{\rm e}/m_{\rm p}$ 
can be used as a probe to search for the hidden scalar fields.

Measurements of this ratio employ a variety of atomic and molecular transitions which 
have a different sensitivity to small changes in $\mu$, or in the fine structure constant $\alpha = e^2/\hbar c$,
or in a combination of $\mu$ and $\alpha$, $F = \alpha^2/\mu$. However, Higgs-like couplings assume that 
the variations of $\alpha$~-- if any~-- are typically much smaller than those of $\mu$ (e.g., Yoo \& Scherrer 2003), 
so that in the first approximation $\alpha$ can be considered as a constant. If relative offsets between different 
molecular frequencies induced by the alleged changes in $\mu$ are large enough, they can be measured
by direct spectroscopic methods in optical and radio bands (Kozlov \& Levshakov 2013; Ubachs 2018).

Up to now, the most accurate probes of $\mu$ were obtained in Galactic molecular clouds distributed in the MW disk 
where the Galactic gravitational potential is the same as in the Solar system and
$\rho_{\rm b}/\rho_{\scriptscriptstyle{\rm DM}} \sim 10$. The upper limit ($1\sigma$) on the fractional change in $\mu$, 
$\Delta\mu/\mu = (\mu_{\rm obs} - \mu_{\rm lab})/\mu_{\rm lab}$, for the MW disk clouds was found to be
$|\Delta \mu/\mu| < 0.9\times10^{-8}$  (Levshakov \etal\ 2013). Employing methanol (CH$_3$OH) absorption-line transitions, 
Kanekar \etal\ (2015) obtained $|\Delta \mu/\mu| < 0.6\times10^{-7}$ for a molecular cloud in a distant galaxy 
at $z = 0.89$ which is a high-redshift analogue to the MW, i.e., the total mass balance in this galaxy
is probably dominated by baryonic matter. On the other hand, for an object in the Large Magellanic Cloud 
located at the galactocentric distance where $\rho_{\rm b}/\rho_{\scriptscriptstyle{\rm DM}} < 1$ we deduced
$\Delta \mu/\mu = (1.7 \pm 0.7)\times10^{-7}$ (Levshakov \etal\ 2019). 
Similarly, a value of $\Delta \mu/\mu = (3.5 \pm 1.2)\times10^{-7}$ was reported by Kanekar (2011) for
a faint dwarf galaxy at $z = 0.69$ where the dark matter may prevail. However, the last two values may be 
affected by non evaluated systematics and require further investigations.

As for the measurements at high-redshifts, previously we estimated the value of
$\Delta F/F = (F_{\rm obs}-F_{\rm lab})/F_{\rm lab}$ for two distant quasars BR~1202--0725 ($z = 4.69$) 
and J1148+5251 ($z = 6.42$) where $|\Delta F/F| < 1.5\times10^{-4}$ (Levshakov \etal\ 2008), and for a 
lensed galaxy HLSJ091828.6+514223 at $z = 5.24$ where $|\Delta F/F| < 1.5\times10^{-5}$ (Levshakov \etal\ 2012).
Later on, the result towards J1148+5251 was slightly improved to $|\Delta F/F| < 4\times10^{-5}$ (Levshakov \etal\ 2017).

Here we evaluate two additional limits on $\mu$-variation using published IRAM/NOEMA spectra of the most 
distant gravitationally lensed quasar J0439+1634 at $z=6.519$ (Yang J. \etal\ 2019; herein referred to as Y19) and
new ALMA observations of the quasar J2310+1855 at $z=6.003$ (Li \etal\ 2020; herein referred to as L20). 
Combining all points ranging between $z = 6.0$ and $z = 6.5$, we obtain the most stringent limit on $\mu$ variations 
at the end of the epoch of reionization.

\section{Data and method}
\label{Sec2}

The radial velocity offset, $\Delta V = V_{\rm rot} - V_{\rm fs}$, between low-lying rotational lines of carbon monoxide
and atomic far infrared (FIR) fine-structure lines, being interpreted in terms of the fractional change in the 
quotient $F = \alpha^2/\mu$, or in the product $F = \tilde{\mu}\alpha^2$ gives (Levshakov \etal\ 2008):
\begin{equation}
{\Delta V}/{c} = \Delta F/F  = 2\Delta\alpha/\alpha - \Delta\mu/\mu = \Delta z/(1 + \bar{z}),
\label{Equ1}
\end{equation}
or
\begin{equation}
{\Delta V}/{c} = 2\Delta\alpha/\alpha + \Delta\tilde{\mu}/\tilde{\mu} = \Delta z/(1 + \bar{z}).
\label{Equ1a}
\end{equation}
Here $\tilde{\mu} = \mu^{-1}$, $c$ is the speed of light, $\Delta z = z_{\rm rot} - z_{\rm fs}$
is the redshift difference between the rotational and fine-structure lines, $\bar{z}$ is their mean redshift,
and $z_{\rm rot}$, $z_{\rm fs}$ are related to the observed and laboratory frequencies, $\nu_{\rm obs}$ and 
$\nu_{\rm lab}$, via
\begin{equation}
z = \nu_{\rm lab}/\nu_{\rm obs} - 1.
\label{Equ2}
\end{equation}
As already mentioned above, Higgs-like couplings do not change $\alpha$. Therefore, in this paper we will assume 
that $\alpha$ is kept fixed. In this approximation, we have
\begin{equation}
\Delta F/F \approx -\Delta\mu/\mu .
\label{Equ2a}
\end{equation}

Velocity offsets \dV\ measured in astrophysical objects usually include also random shifts caused by the
heterogeneous spatial distribution of different species which may not trace each other exactly. On the other hand,
in the present case we analyze integrated emission over the whole surface of a distant galaxy which means that 
such random shifts are already to a great extent averaged. Nevertheless, we define \dV\ as a sum of two
components~--- $\Delta V_F$ due to $F$-variations, and $\Delta V_D$ due to random kinematic effects
(the so-called Doppler noise):
\begin{equation}
\Delta V = \Delta V_F + \Delta V_D.
\label{Equ3}
\end{equation}
The Doppler noise is supposed to be normally distributed with a zero mean and a finite variance, and
the signal $\Delta V_F$ can be estimated statistically by averaging over a data sample:
\begin{equation}
\langle \Delta V \rangle = \langle \Delta V_F \rangle , \;\;\;
{\rm Var}(\Delta V) = {\rm Var}(\Delta V_F) + {\rm Var}(\Delta V_D).
\label{Equ4}
\end{equation}

For a single system, the dispersion of the Doppler noise can be estimated from the comparison of the velocity offsets
between spectral lines of similar species which are linked to each other by a certain physical conditions. 
In our case these are  \CI, \CII, and CO lines. Low-lying rotational lines of carbon monoxide and FIR fine-structure lines
of atomic carbon trace neutral gas which is well shielded from the ionizing radiation (Hollenbach \& Tielens 1999) and, 
as a result, in molecular clouds their profiles are similar (Okada \etal\ 2019). As for \CII, its FIR fine-structure 
emission is usually enhanced at the edges of molecular clouds in the photodissociation regions (PDRs). However,
diffuse gas from the H~{\sc ii} regions can also contribute to some extent ($\la 30$\%, e.g., Kaufman \etal\ 1999) 
to the intensity of the `PDR' lines, being integrated over the surface distribution of the
\CI/\CII/CO emitting gas of a galaxy. As a result, \CII\ lines may have a slightly wider profiles than \CI\ and CO lines
if H~{\sc ii} regions occupy considerable volume of the observed galaxy.

In the present paper, we analyze radial velocity offsets between the rotational lines of CO(5-4) 576.26793050(5) GHz,
CO(6-5) 691.4730763(5) GHz, CO(7-6) 806.65180600(50) GHz, CO(8-7) 921.79970000(50) GHz, CO(9-8) 1036.91239300(50) GHz 
(Endres \etal\ 2016), and the FIR fine-structure lines of \CI\,${}^3{\rm P}_2 \rightarrow {}^3{\rm P}_1$ 809.341970(17) GHz 
(Haris \& Kramida 2017), and \CII\, $^2{\rm P}_{3/2} \rightarrow {}^2{\rm P}_{1/2}$  1900.5369(13) GHz (Cooksy \etal\ 1986).
These lines were observed in emission towards two distant quasars: J0439+1634 at $z = 6.519$ (Y19), and J2310+1855 
at $z = 6.003$ (L20). Among them, only \CI, \CII, CO(6-5), CO(7-6) and CO(9-8) demonstrate simple symmetric profiles
and were taken with the same spectral resolution, $\Delta_{\rm ch} \simeq 50$ \kms\ per channel.

The emission lines of CO(6-5), CO(7-6) and \CI\ were detected in a single setting with receiver 1 (3 mm)
and the angular resolution of $\theta = 6.\!''2 \times 3.\!''2$. The CO(9-8) line was observed with receiver 2 (2 mm) 
and $\theta = 3.\!''1 \times 2.\!''4$, whereas observations of \CII\ were performed with receiver 3 (1 mm) 
in two settings $\theta_1 = 4.\!''9 \times 1.\!''9$, and $\theta_2 = 1.\!''2 \times 0.\!''7$.
In spite of different angular resolutions, all the IRAM/NOEMA detections are spatially unresolved.

However, J0439+1634 is a compact lensed quasar with a maximum image separation $\theta \sim 0.\!''2$
(Fan \etal\ 2019), and this fact may affect line profiles due to differential lensing, i.e., the lensing pattern 
of distant objects may cause complex velocity profiles of emission lines in the spatially unresolved images
in case of different lensing magnifications of emitting regions (e.g., Rivera \etal\ 2019;  Yang C. \etal\ 2019).
This may modify the offset between the relative positions of spectral lines.

A combination of \CI/\CII/CO lines,~--- the so-called Fine Structure Method (FSM),~--- has already been used in differential
measurements of the fundamental physical constants in a wide redshift range~--- from $z \simeq 0$ up to $z \simeq 7$
(Levshakov \etal\ 2008, 2010, 2012, 2017, 2019; Curran \etal\ 2011; Wei\ss\ \etal\ 2012). To note is also that the 
FSM provides, at least, a factor of 30 more sensitive probe of the variability of physical constants than traditionally 
used UV and optical resonance lines of atoms and ions (Levshakov \etal\ 2008; Kozlov \etal\ 2008).

The accurate determination of the centroid $(V_0)$ and the full width at half maximum (FWHM) of a spectral line
is crucial for high-precision measurements. If the line profile is a Gaussian then the statistical uncertainty
of centroid is given by (e.g., Landman \etal\ 1982):
\begin{equation}
\sigma_0 \simeq \frac{0.7 \Delta_{\rm ch}}{\rm SN} \sqrt{n},
\label{Equ5}
\end{equation}
where $\Delta_{\rm ch}$ is the channel width, SN~--- the signal-to-noise ratio, and
$n = {\rm FWHM}/\Delta_{\rm ch}$~--- the line width in units of channels.

Apart from the statistical uncertainty $\sigma_0$, there are systematic errors related to instrumental effects 
such as LSR corrections, $\sigma_{\scriptscriptstyle {\rm LSR}}$, and uncertainties in the rest frame frequencies, 
$\sigma_{\rm rest}$. For instance, a long periodic trend in the LSR corrections due to the effect of Jupiter is
$\sigma_{\scriptscriptstyle \rm LSR} \simeq 12$ \ms, and the largest uncertainty of the rest frame frequency
of $\sigma_{\rm rest} = 200$ \ms\ belongs to \CII\ (Cooksy \etal\ 1986). Both of them is, however, much smaller 
than the channel width $\Delta_{\rm ch} \simeq 50$ \kms\ of the spectral data in question.
If, for a moment, we ignore these systematic errors, $\sigma_{\scriptscriptstyle \rm LSR}$ and $\sigma_{\rm rest}$,
then for a moderate quality data with SN~$\sim 10-20$ and $n \sim 5-6$ the relative statistical error, 
$\delta_0 = \sigma_0/\Delta_{\rm ch}$, is about $1/10th$ of the channel width. On the other hand, for a high quality 
data with SN~$> 100$ the centroid limiting accuracy will be restricted by the systematic errors which for the case of 
the \CII/CO pairs provides a limiting accuracy of $\sigma(\Delta F/F)_{\rm lim} \simeq 7\times10^{-7}$,
whereas for the \CI/CO pairs it is $\simeq 2\times10^{-8}$.

Before processing, we subtracted baselines from each spectrum.  The baseline was defined from the regression analysis of
the mean signals from spectral intervals without emission lines and/or noise spikes. In each such interval the {\it rms} 
noise level was determined as well. Since individual {\it rms} uncertainties were of the same order of magnitude, 
their mean value was assigned to the whole spectrum.

\section{Results}
\label{Sec3}

\subsection{Constraints on $\mu$-variation at $z=6.519$}
\label{Sec3_1}

\begin{figure*}
\vspace*{-1.5cm}
\hspace*{0.0cm}\includegraphics[width=16.0cm]{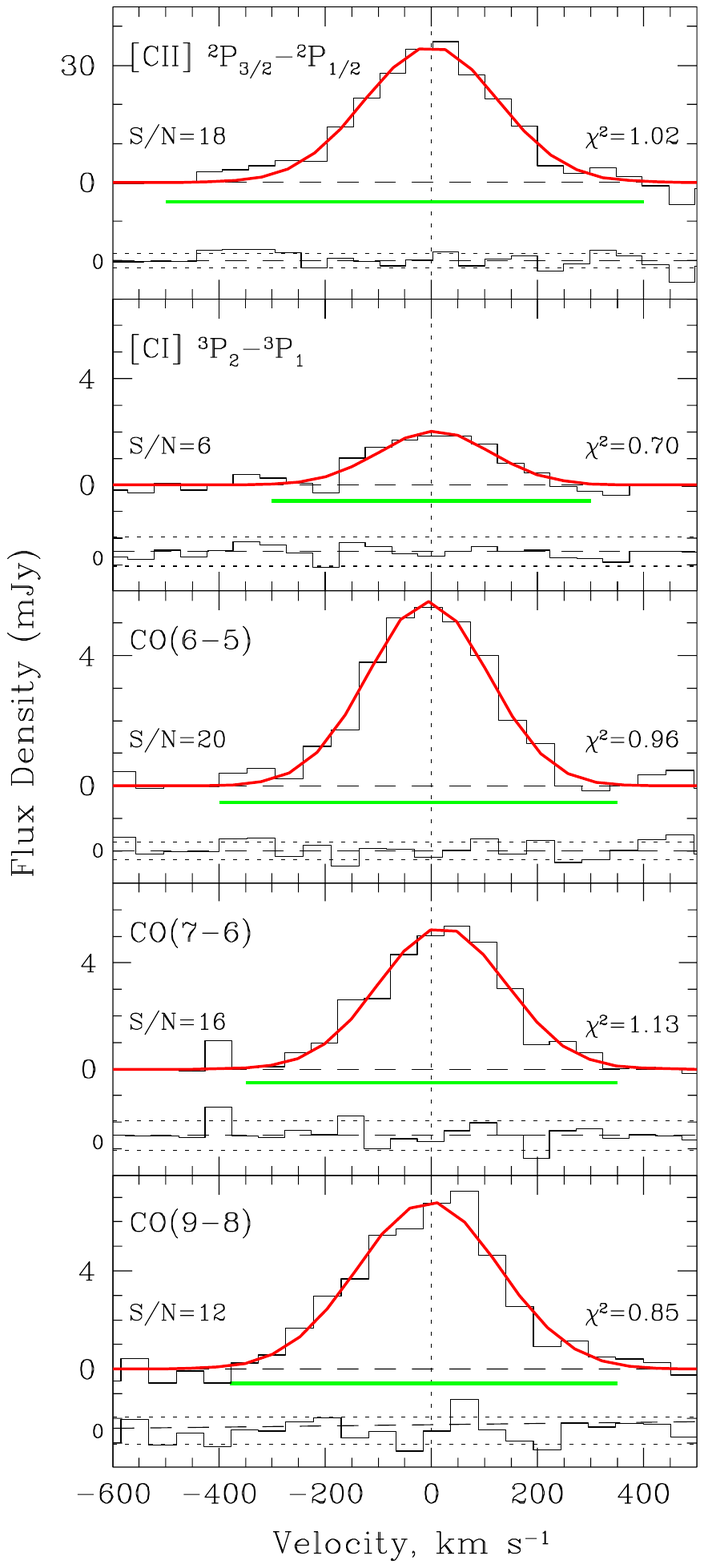}
\vspace*{-3.5cm}
\caption{Black histograms are the baseline subtracted emission lines 
detected by Yang J. \etal\ (2019) towards the quasar J0439+1634.   
The velocity axis is relative to the redshift of \CII\ given in Table~\ref{T1}.
The red continuous lines show the model profiles.
The horizontal green bars specify the velocity range used in the line fitting procedure.
The residuals are plotted by the lower black histogram (arbitrarily offset for clarity).
Two horizontal dotted lines are the mean $\pm 1\sigma$ noise level.
The vertical dotted line marks the \CII\ centroid given to indicate small velocity offsets for other emission lines.
The signal-to-noise ratio (S/N) per bin at the line peak is depicted in each panel.
}
 \label{Fg1}
\end{figure*}

IRAM/NOEMA spectra of J0439+1634 exhibit four rotational transitions of CO:
$J = 6\rightarrow5, 7\rightarrow6, 9\rightarrow8$, and $10\rightarrow9$.
We do not use the last one since it is blended with the H$_2$O 3$_{1,2}$--2$_{2,1}$
emission line. The selected CO lines together with \CII\ and \CI\ 
are shown by black histograms in Fig.~\ref{Fg1}. Each individual line was fitted by 
a single-component Gaussian model (shown by red). The residual uncertainties are plotted 
by the lower black histograms. All emission line profiles are well described by
the Gaussian model with the minimum values of $\chi^2$ per degrees of freedom of about 1.

The derived model parameters are listed in Table~\ref{T1}.
The strongest line is \CII\ with a line flux ${\cal F}_{\scriptscriptstyle \rm [CII]}$ 
more than 5 times exceeding the other line fluxes. It was also observed with a high signal-to-noise ratio. 
Since the measured redshift of $z_{\scriptscriptstyle \rm [CII]} = 6.51877(11)$ is in good
agreement with the previous result $z_{\scriptscriptstyle \rm [CII]} = 6.5188(1)$ of Y19, 
we used \CII\ as a reference line for the velocity scale shown in Fig.~\ref{Fg1}.

\begin{table*}
\centering
\caption{Derived parameters of lines towards the quasar J0439+1634.
$^a$Calculated using rest frame frequencies of 
1900.5369(13) GHz for \CII\ $^2{\rm P}_{3/2} - ^2{\rm P}_{1/2}$ from Cooksy \etal\ (1986),
809.341970(17) GHz for \CI\ $^3{\rm P}_2 - ^3{\rm P}_1$ from Haris \& Kramida (2017),
691.4730763(5) GHz for CO(6-5),
806.65180600(50) GHz for CO(7-6), and
1036.9123930(50) GHz for CO(9-8)
from Endres \etal\ (2016).
Given in parentheses are statistical errors (1$\sigma$) in the last digits.
}
\label{T1}
\begin{tabular}{l r@.l r@.l c r@.l r@.l }
\hline
\multicolumn{1}{c}{Transition} & 
\multicolumn{2}{c}{$\nu_{\rm obs}$} & \multicolumn{2}{c}{FWHM} & FWHM &
\multicolumn{2}{c}{$\cal F$} & \multicolumn{2}{c}{$z^a$}  \\
 & \multicolumn{2}{c}{(GHz)} & \multicolumn{2}{c}{(GHz)} & (\kms) &
\multicolumn{2}{c}{(Jy~\kms)} \\
\hline
\CII     & 252&7725(38) & 0&249(12) & 295(14) & 10&96(5)  & 6&51877(11) \\
\CI      & 107&6406(49) & 0&090(12) & 251(33) &  0&54(15) & 6&51893(34) \\
CO(6-5) & 91&9682(13) & 0&082(4) & 267(13) & 1&61(8) & 6&51861(11) \\
CO(7-6) & 107&2778(18) & 0&101(5) & 282(14) & 1&60(8) & 6&51928(13) \\
CO(9-8) & 137&9120(34) & 0&141(11) & 315(24) & 2&28(15) & 6&51865(19) \\
\hline
\end{tabular}
\end{table*}

In spite of a larger line width of \CII\ than that of \CI\ (FWHN = 294 \kms\ vs. 251 \kms),
both values match each other within the measured $1\sigma$ uncertainty intervals. 
As for the widths of the CO lines, they show a tendency of increasing FWHM with increasing $J$:
the cold gas ($T_{\rm kin} \la 100$~K) tracers CO(6-5) and CO(7-6) have a compatible line width with that of \CI,
whereas CO(9-8) is broader~--- the value of FWHM$_{\scriptscriptstyle \rm CO(9-8)}$ is shifted towards the width of  
FWHM$_{\scriptscriptstyle \rm [CII]}$. Since at the galactic scales the observed line profiles are formed by the global
velocity field, it indicates that the contribution of warm gas ($T_{\rm kin} \sim 200$~K)
to emissivity of \CII\ and CO(9-8) may be essential for the quasar host galaxy J0439+1634.
The observed similarity of line profiles implies that the warm and cold gas are well mixed  
throughout the galactic disk.

As mentioned above, the observed profile of \CII\ is well fitted by a single-component Gaussian model.
However, at $V \simeq -350$ \kms\ in the residual uncertainties (Fig.~\ref{Fg1}) there is an excess of 
four channel widths which may indicate the presence  of a weak blueshifted subcomponent. We note that 
a blueshift of $310\pm120$ \kms\ between the main component of \CII\  and the optical \MgII\ 
($\lambda_0 = 2798$ \AA) emission line from the near-infrared spectrum of J0439+1634 (Fan \etal\ 2019)
was previously revealed in Y19. 

The dispersion of the measured redshifts is illustrated in Fig.~\ref{Fg2}, left panel.
The largest deviation from the reference point $z_{\scriptscriptstyle \rm [CII]}$ is
observed for CO(7-6) with the velocity offset $\Delta V = 20.3$ \kms, which is about one half of the channel size
against an expected $\simeq 0.1\Delta_{\rm ch}$, as in the case of \CI, CO(6-5), and CO(9-8). 
The reason for such a large velocity offset is unclear and probably due to some systematic effects
because CO $J = 7 \rightarrow 6$  is an intermediate transition between $J = 6 \rightarrow 5$ and
$J = 9 \rightarrow 8$, and, hence, its apparent shift cannot be explained as a result of varying  physical
conditions in emitting regions traced by different CO lines. 

To estimate an upper limit on \dmm, at first we average separately the redshifts of the fine structure lines \CI\ and \CII,
and the CO rotational transitions listed in Table~\ref{T1}. It gives $\langle z \rangle_{\rm fs} = 6.51885(18)$ and
$\langle z \rangle_{\rm rot} = 6.51885(9)$. Then we calculated \dmm~= $(\langle z \rangle_{\rm rot} -
\langle z \rangle_{\rm fs})/(1 + \bar{z}) = (0.0 \pm 2.7)\times10^{-5}$ ($1\sigma$).

\subsection{Constraints on $\mu$-variation at $z=6.003$}
\label{Sec3_2}

Here we tighten our constraint on \dmm\ at $z = 6.519$ towards the quasar J2310+1855 at $z=6.003$
using the updated values of velocity offsets between four high-$J$ rotational transitions of CO 
($J = 5 \rightarrow 4$,  $J = 6 \rightarrow 5$,  $J = 8 \rightarrow 7$, and $J = 9 \rightarrow 8$)
and the fine-structure line of \CII\ $^2$P$_{3/2} \rightarrow ^2$P$_{1/2}$. 

In our previous analysis of this quasar (Levshakov \etal\ 2017), we used the CO(6-5) line from  
PdBI observations with angular resolution $\theta = 5.\!''4 \times 3.\!''9$,
and the \CII\ line from ALMA observations with $\theta = 0.\!''72 \times 0.\!''51$ (Wang \etal\ 2013).
The redshift of $z_{\scriptscriptstyle \rm CO(6-5)} = 6.0025(7)$ 
and $z_{\scriptscriptstyle \rm [CII]} = 6.0031(2)$ gave an upper limit on |\dmm| $< 10^{-4}$.
Additional observations of J2310+1855 in CO lines with ALMA (L20; Feruglio \etal\ 2018)
and IRAM/NOEMA (L20) allow us to set a more stringent limit on \dmm. 

The detected molecular lines are listed in Table~2 of L20. Carbon monoxide emission was observed in 
rotational transitions from $J = 2 \rightarrow 1$ up to $J = 13 \rightarrow 12$. 
From this dataset we selected  CO lines obtained with the highest angular resolution and with 
redshifts measured with errors  $\sigma_z \leq 0.0004$ (Table~\ref{T2}). 
The reported line widths are distributed around the mean 
$\langle {\rm FWHM} \rangle_{\rm CO} = 392$ \kms\ with standard deviation $\sigma_{\rm FWHM} = 25$ \kms. 
The width of \CII\ is in good agreement with CO lines, i.e., they all are, most probably, co-spatially distributed.

\begin{figure*}
\hspace*{0.5cm}\includegraphics[width=12.0cm]{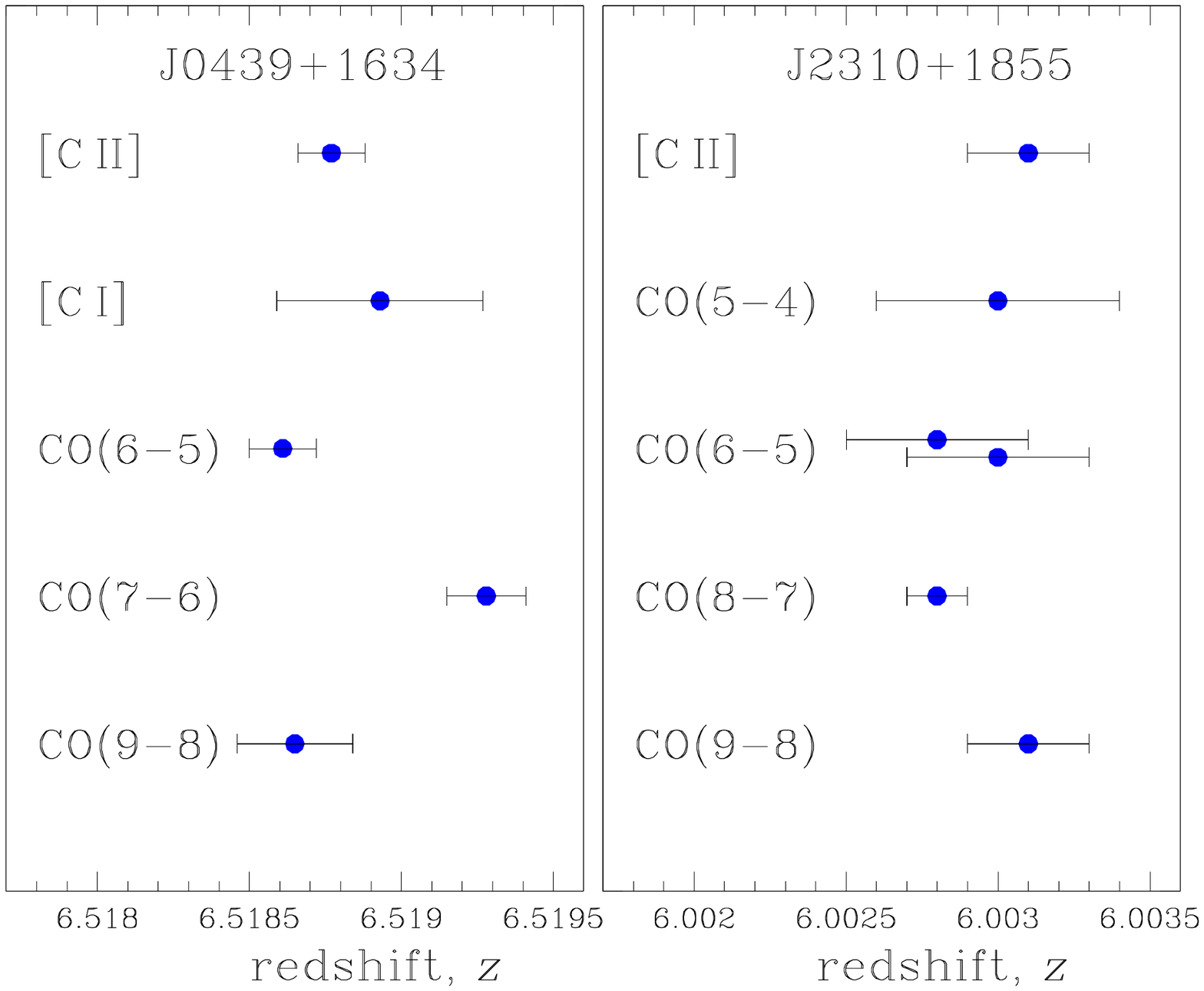}
\vspace*{-3.5cm}
\caption{Measured redshifts of spectral lines towards J0439+1634 and J2310+1855 
and their $1\sigma$ statistical errors listed in Table~\ref{T1} and \ref{T2}, respectively.
The two observations of CO(6-5) towards J2310+1855 were obtained at different telescopes
which are indicated in Table~\ref{T2}.
}
 \label{Fg2}
\end{figure*}

The measured redshifts with their $1\sigma$ error bars are shown in Fig.~\ref{Fg2}, right panel.
Averaging five CO redshifts from Table~\ref{T2} gives $\langle z \rangle_{\rm rot} = 6.00294(13)$. 
With $z_{\rm fs} = z_{\scriptscriptstyle \rm [CII]} = 6.0031(2)$
the updated result at $z = 6.003$ yields \dmm~= $(2.3\pm3.4)\times10^{-5}$ ($1\sigma$).

\begin{table*}
\centering
\caption{Selected parameters of lines towards J2310+1855 (Li \etal\ 2020). 
References: ${}^1$Wang \etal\ 2013; ${}^2$Li \etal\ 2020; ${}^3$Feruglio \etal\ 2018. 
Given in parentheses are statistical errors (1$\sigma$) in the last digits.
}
\label{T2}
\begin{tabular}{l c c c c c} 
\hline
\multicolumn{1}{c}{Transition} & 
\multicolumn{1}{c}{$z$} & \multicolumn{1}{c}{FWHM} & 
\multicolumn{1}{c}{$\Delta_{\rm ch}$} & $\theta$ & Facilities \\
 &  & (\kms) & \multicolumn{1}{c}{(\kms)} & (arcsec) \\
\hline
\CII    & 6.0031(2) & 393(21) & 18 &  0.72$\times$0.51 & ALMA$^1$ \\

CO(5-4) & 6.0030(4) & 409(44) & 60 & 1.67$\times$1.37 & NOEMA$^2$ \\

CO(6-5) & 6.0028(3) & 361(9)  & 23.7 & 0.6$\times$0.4   &  ALMA$^3$ \\

CO(6-5) & 6.0030(3) & 422(20) & 60 & 1.42$\times$1.19 &  NOEMA$^2$ \\

CO(8-7) & 6.0028(1) & 390(15) & 36 & 0.79$\times$0.75 & ALMA$^2$ \\  

CO(9-8) & 6.0031(2) & 376(18) & 32 &  0.77$\times$0.63 & ALMA$^2$ \\
\hline
\end{tabular}
\end{table*}

\section{Possible $z$-dependence of $\mu$ }
\label{Sec4}

If we assume that the result of Hart \& Chluba (2020), 
$m_{{\rm e},z} = (1.0190\pm0.0055)m_{\rm e,0}$, is real at the recombination epoch, 
then in combination with constraints on \dmm\ at lower redshifts, $z < 7$,
this implies that there should be a redshift dependence of non-standard scalar field(s) coupled to ordinary matter.
All available measurements of \dmm\ from the range $z \in [0,1100]$ are plotted by dots with error bars in Fig.~\ref{Fg3},
while the detailed information on individual data points is given in Appendix. Two inserts with expanded $Y$-scales 
are used to make the error bars visible in the indicated redshift intervals. 

To approximate the $z$-dependence of $\mu$ we employ a simple power law
\begin{equation}
\frac{\Delta\mu}{\mu} = k_\mu (1+z)^p\; ,
\label{Equ6}
\end{equation}
with $k_\mu$ and $p$ being the model parameters. A similar form was used in Hart \& Chluba (2018, 2020).

The regression analysis yields $k_\mu = (1.6 \pm 0.3)\times10^{-8}$ and $p = 2.00 \pm 0.03$ ($1\sigma$)
with the corresponding regression curve shown by red in Fig.~\ref{Fg3}.
We note that the null result for the power index $p$ reported by Hart \& Chluba (2020), 
$p = (0.7 \pm 3.1)\times 10^{-3}$, is due to the fact that their calculations were restricted 
by the epoch of recombination only, $\Delta z \simeq 200$ at $z = 1100$.

We also tried an exponential function depending on two parameters $k'_\mu$ and $p'$: 
\begin{equation}
\frac{\Delta\mu}{\mu} = k'_\mu \exp( p' z)\; .
\label{Equ6a}
\end{equation}
The corresponding curve for $k'_\mu = 1.7\times10^{-8}$ and $p' = 0.013$ is shown by the dashed grey curve 
in Fig.~\ref{Fg3}.  It is clearly seen that the points available do not allow to decide between the power law 
and exponential functions. However, both of them lead to a positive coefficient  $k_\mu$ ($k'_\mu$) at $z \rightarrow 0$
if the effective electron rest mass at the epoch of recombination differs from its terrestrial value.

The question arises how the obtained redshift dependence of \dmm\ with $k_\mu > 0$ can be verified?

It is obvious that the recombination point, which is crucial for the present analysis,
should be confirmed in further studies of the cosmic microwave background anisotropies. 
We note that the fractional changes in $m_{\rm e}$ of $\sim$~1-2\% 
do not contradict the results of primordial nucleosynthesis ($z \sim 10^8$) which show that 
abundances of the light nuclei are not sensitive to the changes in $ m_{\rm e}$ within $\pm10$\% 
(e.g., Coc \etal\ 2007; Uzan 2011).

At $z \sim 17$, the central redshift of the observed H\,{\sc i} 21-cm absorption, 
the expected value of \dmm\ is $5\times10^{-6}$ [Eq.(\ref{Equ6})] or $2\times10^{-8}$ [Eq.(\ref{Equ6a})]. 
Measurements at such level of sensitivity can be provided solely by radio spectroscopy of molecular and atomic transitions.  
However, presently neither molecular nor atomic transitions have been detected above $z \sim 10$
albeit galaxies with redshifts $z > 10$ are expected to be observed with the Atacama Large Millimeter/submillimeter Array 
(de Blok \etal\ 2016) and the James Webb Space Telescope (Behroozi \etal\ 2020).

For the epoch of reionization ($6 < z < 10$) the expected value of \dmm\ is less than $10^{-6}$.
To date, the best estimate of the fractional changes in $\mu$ at $z > 6$ is $|\Delta \mu/\mu| < 10^{-5}$, i.e.,
the measurement accuracy should be improved by an order of magnitude.  According to Eq.~(\ref{Equ5}), this level 
can be achieved if the exposure time will be increased by $\sim 100$ times at the fixed values of the channel width
$\Delta_{\rm ch} \simeq 50$ \kms\ and the line width FWHM~$\simeq 200$ \kms. With the existing observing facilities 
this seems to be problematic if one deals with \CI/\CII/CO transitions. 

Another way to reach an accuracy $\sim 10^{-6}$ is to use spectral lines
with higher sensitivity coefficients, $Q_\mu$,  to changes in $\mu$ [see Eq.(\ref{Equ1A}) in Appendix].
For instance, the $1_1 - 2_2 E$ transition of the methanol isotopologue ${}^{12}$CD$_3{}^{16}$OH at
a rest frequency of $\nu \approx 1.2$ GHz has $Q_\mu = -330$ (Jansen \etal\ 2011),
the sensitivity coefficients of the $\Lambda$-doublet hyperfine components of the
${}^2\Pi_{1/2} J=9/2$ state of OH from the interval $\Delta\nu = 89-193$ MHz 
are ranging between $Q_\mu = 212$ and 460 (Kozlov 2009).
The sensitivities of these transitions to variations in $\mu$ are more than
two orders of magnitude larger than $Q_\mu = 1$ 
used in the current studies with \CI/CO and \CII/CO pairs.
These and other high sensitivity molecular transitions are planned to be observed 
with the Five-hundred-meter Aperture Spherical radio Telescope (Chen \etal\ 2019).

Observations of local objects ($z \sim 0$) have an important specificity that
their angular sizes are larger than the aperture of a telescope.
This makes possible the scanning of the local objects across their surfaces.
Then spatial fluctuations of $\mu$ can be expected
if some screening mechanism predicted in a number of theories is indeed at play.
The currently available data show that the upper limit on
the amplitude of such fluctuations is $k_\mu \la 10^{-8}$.
Thus, to detect the signal the spectral line positions should be measured with 
uncertainties of less or about 0.01 \kms. Such an accuracy can already be achieved at the 
existing observing facilities.

\section{Summary} 
\label{Sec5}

Our main results are as follows:
\begin{itemize}
\item[$\bullet$]
Using a combination of the \CI\ and \CII\ fine structure lines together with
CO(6-5), (7-6), and (9-8) rotational transitions observed towards the quasar J0439+1634 by Yang J. \etal\ (2019),
we set a limit on \dmm~$= (0.0 \pm 2.7)\times10^{-5}$ at $z = 6.519$.
\item[$\bullet$]
A combination of the \CII\ fine structure line with 
CO(5-4), (6-5), (8-7), and (9-8) rotational transitions
from the spectrum of J2310+1855 (Li \etal\ 2020) yields
\dmm~$= (2.3 \pm 3.4)\times10^{-5}$ at $z = 6.003$.
\item[$\bullet$]
Two values of \dmm\ at $z = 6.519$ and $z = 6.003$, being combined with
\dmm~$= (-1.3 \pm 7.9)\times 10^{-5}$ at $z = 6.419$ towards the quasar  
J1148+5251 (Levshakov \etal\ 2017), give a mean value of
$\langle \Delta\mu/\mu \rangle = (0.7 \pm 1.2)\times 10^{-5}$ at
$\bar{z} = 6.3$ which is the most stringent up-to-date limit
on the fractional changes in $\mu$ at the epoch of reionization.
\item[$\bullet$]
Exploiting the value of 
$m_{{\rm e},z} = (1.0190\pm0.0055)m_{\rm e,0}$ at $z = 1100$
from Hart \& Chluba (2020) and all available data on \dmm, we obtain
a functional $z$-dependence of $\mu$ in the form
\dmm~$= k_\mu (1+z)^p$ with $k_\mu = (1.6 \pm 0.3)\times 10^{-8}$ and
$p = 2.00 \pm 0.03$.  Possible ways to verify 
this dependence by further observations are discussed.
\end{itemize}

\section*{Acknowledgments}
We would like to thank our referee Wim Ubachs for useful comments and suggestions.
This work was supported by the Russian Science
Foundation under grant No.~19-12-00157.

\section*{Data Availability}

Based on the published spectra obtained at
the IRAM NOEMA Interferometer (project S18DO and W18EI)
and at the ALMA Interferometer (project ADS/JAO.ALMA 2015.1.01265.S).

IRAM is supported by INSU/CNRS (France), MPG (Germany), and IGN
(Spain). The National Radio Astronomy Observatory is
a facility of the National Science Foundation operated
under cooperative agreement by Associated Universities, Inc.

ALMA is a partnership of ESO, NSF (USA) and NINS (Japan), together with NRC (Canada), MOST
and ASIAA (Taiwan), and KASI (Republic of Korea), in cooperation with the Republic of Chile.

Spectral data used in this study are publicly available from the 
corresponding papers cited herein.

\begin{figure*}
\hspace*{-2.0cm}\includegraphics[width=13.5cm]{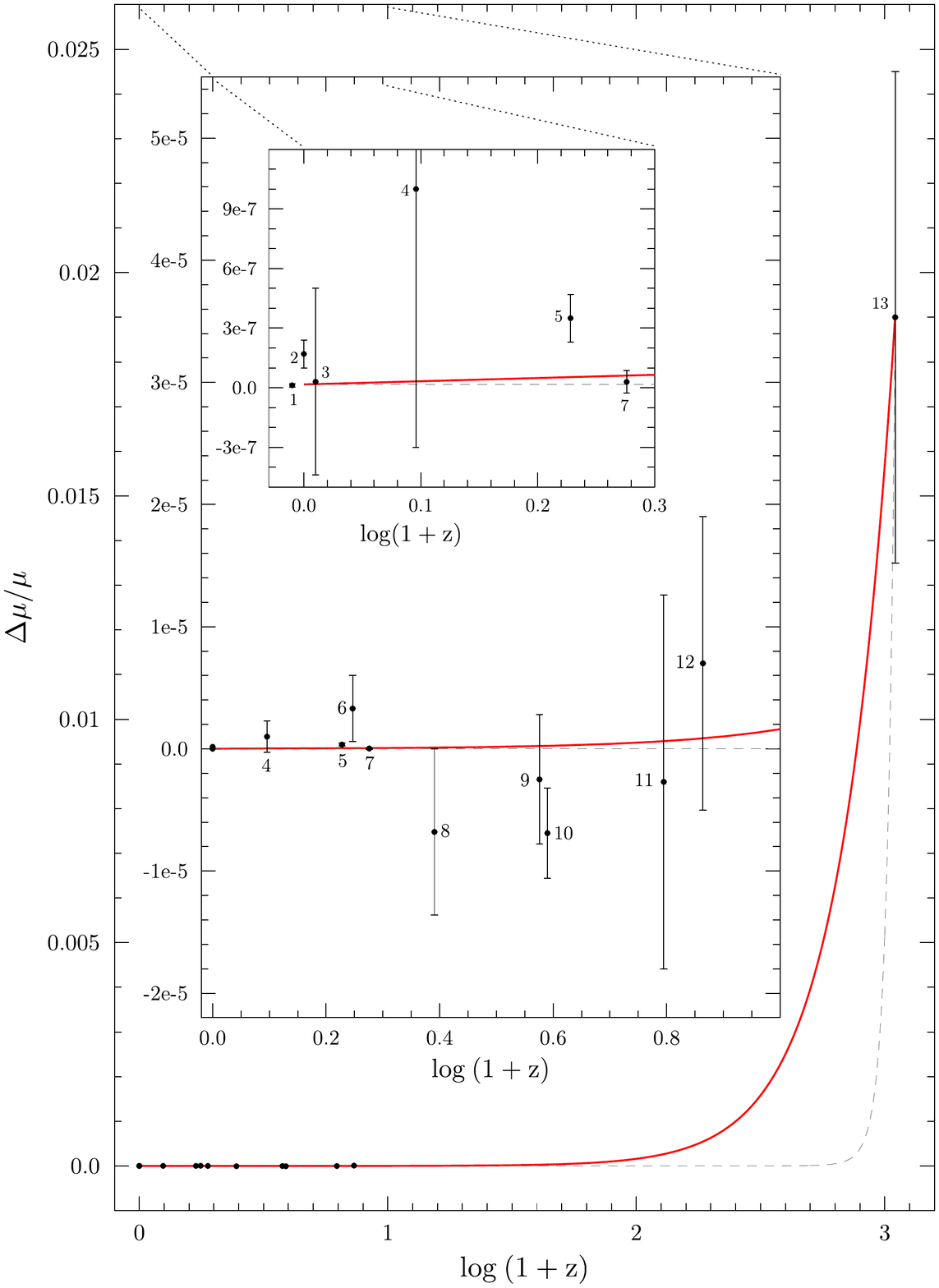}
\vspace*{0.0cm}
\caption{Constraints on the fractional changes in $\mu$ (dots with $1\sigma$ error bards)
as a function of redshift $z$ in units of log$_{10}(1+z)$. Point 13 represents the fractional change 
in the rest electron mass $m_{{\rm e},0}$ at the epoch of recombination ($z \sim 1100$).
Two inserts zoom consequently the corresponding parts of the data sample using
different horizontal and vertical scales. Points 1,2, and 3 at $z = 0$ are slightly shifted with respect 
to each other in order to resolve blending. Shown by red is a two-parameter 
regression curve \dmm~= $k_\mu (1+z)^p$ with $k_\mu = (1.6 \pm 0.3)\times10^{-8}$
and $p = 2.00 \pm 0.03$ ($1\sigma$). The dashed grey curve is an exponential function 
\dmm~= $k'_\mu \exp(p' z)$ with $k'_\mu = 1.7\times10^{-8}$ and $p' = 0.013$.
References for data points: 
1~-- Levshakov \etal\ (2013); 
2~-- Levshakov \etal\ (2019);
3~-- Levshakov \etal\ (2017);
4~-- Kanekar \etal\ (2018);
5~-- Kanekar (2011);
6~-- Kanekar \etal\ (2012);
7~-- Kanekar \etal\ (2015);
8~-- Kanekar \etal\ (2010);
9~-- Ubachs \etal\ (2016, 2019);
10~-- Wie\ss\ \etal\ (2012);
11~-- Levshakov \etal\ (2012);
12~-- this work;
13~-- Hart \& Chluba (2020).
}
 \label{Fg3}
\end{figure*}

\appendix
\section{Description of the data points}

The data points shown in Fig.~\ref{Fg3} were compiled from the following sources.

\bigskip\noindent
{\bf Milky Way disk $(z=0)$}. 
In the Milky Way, high-resolution spectral observations of dark clouds were performed 
in the inversion line of NH$_3$(1,1) and pure rotational lines of other molecules
(the so-called ammonia method) at the Medicina 32-m and the Effelsberg 100-m radio telescopes to
measure the radial velocity offset, $\Delta V = V_{\rm rot} - V_{\rm inv}$,
between the rotational and inversion transitions to calculate 
\begin{equation}
\frac{\Delta\mu}{\mu} = \frac{V_{\rm rot} - V_{\rm inv}}{c(Q_{\rm inv} - Q_{\rm rot})} 
\approx 0.3\frac{\Delta V}{c},
\label{Equ1A}
\end{equation}
where $c$ is the speed of light, and $Q_{\rm inv}, Q_{\rm rot}$ are the corresponding
sensitivity coefficients to changes in $\mu$ (Flambaum \& Kozlov 2007).

In Effelsberg observations, 19 independent offsets of \dV\ 
gave a weighted mean $\langle \Delta V \rangle = 0.003 \pm 0.006$ \kms\ ($1\sigma$)
which constrained the $\mu$-variation at the level of 
\dmm\ = $(0.3 \pm 0.6)\times10^{-8}$.
The Medicina observations of two dark clouds L1521 and L1498 provided respectively
\dmm\ = $(0.1 \pm 2.2)\times10^{-8}$ and
\dmm\ = $(-0.1 \pm 2.3)\times10^{-8}$ (Levshakov \etal\ 2013).

Later on, the dark cloud core L1498 was observed with the IRAM 30-m telescope in
methanol CH$_3$OH lines (Dapr\`a \etal\ 2017), which resulted in 
$\Delta \tilde{\mu}/\tilde{\mu} = (-3.3 \pm 1.9)\times10^{-8}$ or
\dmm\ = $(3.3 \pm 1.9)\times10^{-8}$.
Thus, a combined Medicina and IRAM limit on \dmm\ towards L1498 is 
\dmm\ = $(3.2 \pm 1.5)\times10^{-8}$.
Now, if we add this value to both the Medicina L1521 and Effelsberg 19 clouds measurements, then
the mean constraint on \dmm\ in the MW disk reads
$\langle \Delta\mu/\mu \rangle = (1.2 \pm 0.9)\times10^{-8}$  ($1\sigma$).

\bigskip\noindent
{\bf Magellanic Clouds $(z=0)$}.  Analyzing data of 9 molecular clouds in the Large and Small Magellanic Clouds,
we obtained for a highest resolution spectrum of a target PDR3-NE (LMC) an offset
$\Delta V  = -0.05 \pm 0.02$ \kms\ between the CO(7-6) and \CI\ lines (Levshakov \etal, 2019).   
Being interpreted in terms of $\alpha^2/\mu$ variations, 
this gives $\Delta F/F  = (-1.7 \pm 0.7)\times10^{-7}$, or 
$\Delta \mu/\mu = (1.7 \pm 0.7)\times10^{-7}$ ($1\sigma$) if we assume that
$|\Delta \alpha/\alpha | \ll |\Delta \mu/\mu|$.

\bigskip\noindent
{\bf Triangulum galaxy M33 $(z \approx 0)$}. 46 emitters in the CO(2-1) and \CII\ lines towards M33 show an offset
$\langle \Delta V \rangle = -0.01 \pm 0.14$ \kms\ (Levshakov \etal\ 2017), which corresponds to
$\langle \Delta F/F \rangle = (-0.3 \pm 4.7)\times10^{-7}$,
or $\langle \Delta\mu/\mu \rangle = (0.3 \pm 4.7)\times10^{-7}$ ($1\sigma$).

\bigskip\noindent
{\bf Quasar PKS 1413+135 $(z=0.25)$}. The conjugate satellite OH 18~cm lines at redshift $z=0.247$ observed
in emission (1720 MHz line) and absorption (1612 MHz) towards the BL Lacertae-type quasar PKS 1413+135
yield \dmmt\ = $(-1.0 \pm 1.3)\times10^{-6}$ (Kanekar \etal\ 2018), or
\dmm~= $(1.0 \pm 1.3)\times10^{-6}$ ($1\sigma$).

\bigskip\noindent
{\bf Gravitational lens system B0218+357 $(z=0.69)$}. The ammonia method was applied to 
the inversion (NH$_3$) and rotational (CS, H$_2$CO) absorption lines detected in a lensed galaxy at 
$z = 0.69$ towards B0218+357 (Kanekar 2011).
The measured fractional changes in $\mu$ are limited 
at the level of \dmm~= $(3.5 \pm 1.2)\times10^{-7}$ ($1\sigma$).

\bigskip\noindent
{\bf Quasar J0134--0931 $(z=0.765)$}. 
A comparison between H\,{\sc i} 21-cm and OH 18-cm absorption lines 
sets a limit on $\Delta X/X = (-5.2 \pm 4.3)\times10^{-6}$,
where $X = g_{\rm p}(\tilde{\mu}\alpha^2)^{1.57}$ and $g_{\rm p}$ is the proton g-factor
(Kanekar \etal\ 2012).
Assuming that fractional changes in $g_{\rm p}$ and $\alpha$ are smaller than those in $\mu$,
we obtain
\dmm~= $(3.3 \pm 2.7)\times10^{-6}$ where error includes both the statistical and systematic uncertainties.

\bigskip\noindent
{\bf Gravitational lens system PKS1830--211 $(z=0.89)$}. In Kanekar \etal\ (2015),
the $z = 0.89$ gravitational lens towards PKS1830--211
was observed in methanol CH$_3$OH absorption lines yielding \dmm~= $(0.29 \pm 0.57)\times10^{-7}$
($1\sigma$). 

\bigskip\noindent
{\bf Quasars Q~2337--011 $(z=1.36)$ and Q~0458--020 $(z=1.56)$}. 
Two absorbers at $z=1.36$ and $z=1.56$ towards respectively Q~2337--011 and Q~0458--020
were studied in the H\,{\sc i} 21-cm and C\,{\sc i} $\lambda\lambda$1560, 1657 \AA\ lines
to set constraints on the product $X = g_{\rm p}\tilde{\mu}\alpha^2$,
where $g_{\rm p}$ is the proton g-factor (Kanekar \etal\ 2010).
The mean value of $\Delta X/X = (6.8 \pm 1.0_{\rm stat} \pm 6.7_{\rm sys})\times10^{-6}$
transforms to \dmm~= $(-6.8 \pm 6.8)\times 10^{-6}$ at $\bar{z} = 1.46$ if we assume that 
fractional changes in $g_{\rm p}$ and $\alpha$ are smaller than those in $\mu$.

\bigskip\noindent
{\bf H$_2$ absorption-line systems $(\bar{z} = 2.76)$}. In the interval 
from $z = 2.05$ to $z = 4.22$,  
the $\mu$-variation can be constrained from a sample of nine H$_2$ systems 
selected from Ubachs \etal\ (2016, 2019). We list these measurements of \dmm\
along with their weighted mean value in Table~\ref{Ta1}. 
At the mean redshift $\bar{z} = 2.763$, one finds 
$\langle \Delta\mu/\mu \rangle = (-2.5 \pm 5.3)\times10^{-6}$.

\begin{table*}
\centering
\caption{Selected H$_2$ absorption systems from
${}^1$Ubachs \etal\ 2016, and ${}^2$Ubachs \etal\ 2019. 
}
\label{Ta1}
\begin{tabular}{l c r@.l r@.l c} 
\hline
\multicolumn{1}{c}{Quasar} & 
\multicolumn{1}{c}{$z$(H$_2$)} & \multicolumn{2}{c}{$\Delta\mu/\mu$ } &  
\multicolumn{2}{c}{$\sigma_{\Delta\mu/\mu}$ } & Refs. \\
 & & \multicolumn{2}{c}{$(\times10^{-6})$ } &
\multicolumn{2}{c}{$(\times10^{-6})$} \\
\hline
J2123--0050  & 2.05 & $-7$&6   &  3&5  & 1 \\[-2pt]
HE0027--1836 & 2.40 &    7&6   & 10&2  & 1 \\[-2pt] 
Q2348--011   & 2.43 &    6&8   & 27&8  & 1 \\[-2pt]
Q0405--443   & 2.59 & $-7$&5   &  5&3  & 1 \\[-2pt] 
B0642--5038  & 2.66 & $-9$&44  &  6&09 & 2 \\[-2pt]
J1237+064    & 2.69 &    4&37  &  6&30 & 2 \\[-2pt]
Q0528--250   & 2.81 &    0&5   &  2&7  & 1 \\[-2pt] 
Q0347--383   & 3.02 & $-5$&1   &  4&5  & 1 \\[-2pt]
J1443+2724   & 4.22 &    9&5   &  7&5  & 1 \\[2pt]
\multicolumn{2}{r}{weighted mean:} & $-2$&5  & 5&3 \\
\hline
\end{tabular}
\end{table*}

\bigskip\noindent
{\bf QSO host galaxy RXJ0911.4+0551 $(z=2.79)$}. One of the most attractive target for high-precision \dFF\ 
measurements is the quasar host galaxy RXJ0911.4+0551 at $z = 2.79$ which emits very strong and narrow
CO(7-6) and \CI\ lines (Wei\ss\ \etal\ 2012). The FSM applied to this system yields 
\dFF~= $(6.9 \pm 3.7)\times10^{-6}$, or neglecting contribution from \daa, we have 
\dmm~= $-(6.9 \pm 3.7)\times10^{-6}$ ($1\sigma$).

\bigskip\noindent
{\bf Lensed galaxy HLSJ091828.6+514223 $(z=5.24)$}. An upper limit on
\dFF\ was deduced from observation of CO(7-6) and \CI\ lines towards the lensed galaxy
HLSJ091828.6+514223 at $z = 5.243$ (Levshakov \etal\ 2012),
which corresponds to \dmm~= $(-0.27 \pm 1.53)\times10^{-5}$ ($1\sigma$). 

\bigskip\noindent
{\bf High-redshift \CI/\CII/CO systems $(\bar{z} = 6.3)$}. Two estimates of 
\dmm~= $(0.0\pm2.7)\times10^{-5}$ at $z = 6.519$
and \dmm~= $(2.3\pm3.4)\times10^{-5}$ at $z = 6.003$ from the previous section
can be combined with another high redshift value of \dmm~= $(-1.3\pm7.9)\times10^{-5}$
at $z = 6.419$ towards the quasar J1148+5251  (Levshakov \etal\ 2017).
Being averaged, 
these three estimates provide a weighted mean value of
$\langle \Delta\mu/\mu \rangle = (0.7\pm1.2)\times10^{-5}$ 
at $\bar{z} = 6.3$ in the epoch of reionization.

\bigskip\noindent
{\bf The CMB limit on electron mass changes $(z \simeq 1100)$}. 
In recent analysis of the cosmic microwave background radiation
by Hart \& Chluba (2020), it was shown that the electron rest mass, $m_{\rm e,0}$,
might be slightly increased during the cosmological recombination era at $z \simeq 1100$: 
$\Delta m_{\rm e}/m_{\rm e,0} = (m_{{\rm e},z} - m_{\rm e,0})/m_{\rm e,0} =
0.0190 \pm 0.0055$ ($1\sigma$).

An important point to note for the CMB analysis is
that while only dimensionless numbers are invariant to the adopted system of units,
variation of the dimensional parameter $m_{\rm e}$ 
for the physics of recombination is equivalent after rescaling
to a variation in the dimensionless parameter $\mu$, 
and ``{\it the effect of a variation in $m_{\rm p}$
on recombination is subdominant compared to a variation in $m_{\rm e}$}'' 
(Planck Collaboration \etal\ 2015).  
The binding energy of quarks can be also altered
but its impact on the atomic and molecular frequencies and on the CMB spectrum
is much weaker than the varying electron mass
(Kujat \& Scherrer 2000; Yoo \& Scherrer 2003; Ichikawa \etal\ 2006).

\bsp    
\label{lastpage}
\end{document}